\documentclass[twocolumn]{aastex7}
\pdfoutput=1 %for arXiv submission
\usepackage{appendix}
\usepackage{amsmath,amstext}
\usepackage[T1]{fontenc}
\usepackage{graphicx}
\usepackage[figure,figure*]{hypcap}
\usepackage[export]{adjustbox}
\usepackage{hyperref}
\usepackage{cleveref}
\graphicspath{{./}{figures/}}

\shorttitle{K2-232c: The Hidden Giant}
\shortauthors{Ranshaw et al.}
\newcommand{\bjdtdb}{\ensuremath{\rm {BJD_{TDB}}}}
\newcommand{\feh}{\ensuremath{\left[{\rm Fe}/{\rm H}\right]}}

\newcommand{\teff}{\ensuremath{T_{\rm eff}}}
\newcommand{\teq}{\ensuremath{T_{\rm eq}}}

\newcommand{\ecosw}{\ensuremath{e\cos{\omega_{*}}}}
\newcommand{\esinw}{\ensuremath{e\sin{\omega_{*}}}}

\newcommand{\msun}{\ensuremath{\,M_\Sun}}
\newcommand{\rsun}{\ensuremath{\,R_\Sun}}
\newcommand{\lsun}{\ensuremath{\,L_\Sun}}

\newcommand{\mj}{\ensuremath{\,M_{\rm J}}}

\newcommand{\rj}{\ensuremath{\,R_{\rm J}}}
\newcommand{\re}{\ensuremath{\,R_{\rm \Earth}}\xspace}
\newcommand{\me}{\ensuremath{\,M_{\rm \Earth}}\xspace}
\newcommand{\fave}{\langle F \rangle}
\newcommand{\fluxcgs}{10$^9$ erg s$^{-1}$ cm$^{-2}$}

\newcommand{\kepler}{{\it Kepler}}
\newcommand{\ktwo}{{\it K2}}
\newcommand{\tess}{{\it TESS}}
\newcommand{\gaia}{{\it Gaia}}

\newcommand{\aref}[1]{\hyperref[#1]{Appendix~\ref*{#1}}}
\newcommand{\degrees}{\ensuremath{^{\circ}}}
\usepackage{caption}
\begin{document}

\title{The Discovery of K2-232c: \\
Divergent Formation Histories for Hot and Warm Jupiters Based on Outer Companion Eccentricity}

\author[0000-0003-0941-4181]{Jessica A. Ranshaw}
\affil{Department of Astronomy, Indiana University, Bloomington, IN 47405}
\email{jranshaw@iu.edu}

\author[0000-0002-0376-6365]{Xian-Yu Wang}
\altaffiliation{Sullivan Prize Postdoctoral Fellow}
\affil{Department of Astronomy, Indiana University, Bloomington, IN 47405}
\email{xwa5@iu.edu}

\author[0000-0002-0040-6815]{Jennifer A. Burt}
\affil{Jet Propulsion Laboratory, California Institute of Technology, 4800 Oak Grove Drive, Pasadena, CA 91109}
\email{Jennifer.Burt@jpl.nasa.gov}

\author[0000-0001-9480-8526]{Juan I. Espinoza-Retamal}
\affiliation{Department of Astrophysical Sciences, Princeton University, 4 Ivy Lane, Princeton, NJ 08540, USA}
\affiliation{Instituto de Astrof\'isica, Pontificia Universidad Cat\'olica de Chile, Av. Vicu\~na Mackenna 4860, 782-0436 Macul, Santiago, Chile}
\affiliation{Millennium Institute for Astrophysics, Santiago, Chile}
\email{jiespinozar@uc.cl}

\author[0000-0002-0015-382X]{Brandon T. Radzom}
\affiliation{Department of Astronomy, Indiana University, Bloomington, IN 47405}
\affiliation{Caltech/IPAC-NASA Exoplanet Science Institute, 1200 E. California Boulevard, MC 100-22, Pasadena, CA 91125, USA}
\email{bradzom@iu.edu}

\author[0000-0003-1305-3761]{R. Paul Butler}
\affil{Earth and Planets Laboratory, Carnegie Science, 5241 Broad Branch Road NW, Washington, DC 20015, USA}
\email{pbutler@carnegiescience.edu}

\author[0000-0002-6153-3076]{Bradford P. Holden}
\affil{UCO/Lick Observatories, University of California, Santa Cruz, 95065, USA}
\email{holden@ucolick.org}

\author[0000-0001-7177-7456]{Steven S. Vogt}
\affil{UCO/Lick Observatory, Department of Astronomy and Astrophysics, University of California at Santa Cruz, Santa Cruz, CA 95064}
\email{vogt@ucolick.org}

 \correspondingauthor{Jessica A. Ranshaw}
 \email{jranshaw@iu.edu}

%\author[0000-0002-3253-2621]{Gregory Laughlin}
%\affil{Department of Astronomy, Yale UniCversity, New Haven, CT, 06511, USA}
%\email{greg.laughlin@yale.edu}

\author[0000-0002-7846-6981]{Songhu Wang}
\affil{Department of Astronomy, Indiana University, Bloomington, IN 47405}
\email{sw121@iu.edu}

\begin{abstract}
    Ever since their discovery, hot Jupiters have been one of the most studied types of exoplanets to exist thanks to their significant size, their proximity to their host star, and their significant departure from anything present in our solar system. Yet, the details of their formation and evolution remain unclear, including their connection, if any, to the wider-orbiting warm Jupiter population. In this work, we present the discovery of K2-232c, an eccentric cold Jupiter ($P = 1950 ^{+140}_{-120}$ days, $e=0.352^{+0.095}_{-0.076}$, $M\sin{i} = 5.31^{+0.48}_{-0.45} \mj$) companion in a known warm Jupiter ($P = 11.1684377 \pm 0.0000010$ days, $e=0.245^{+0.023}_{-0.024}$, $M = 0.427^{+0.039}_{-0.036} \mj$) system. Placing this system in context with the literature, we find that cold Jupiter eccentricities are generally higher in hot Jupiter systems as compared to warm Jupiter systems, suggesting the formation of hot Jupiters is more dynamically violent than warm Jupiters. This adds further evidence to the claim that these two populations are independent from one another with cold Jupiters appearing to play a crucial role in shaping the formation of both.  
\end{abstract}

\section{Introduction}

Hot Jupiters ($M$ = 0.3-13$\mj$, $a < 0.1$ AU) are a well characterized population in the field of exoplanet research, yet their origins remain elusive. Two distinct paths can describe a hot Jupiter's formation, one corresponding to dynamically violent processes (e.g., high-eccentricity migration, \citealt{Rasio1996, Weidenschilling1996, Wu2003,Chatterjee2008,WuLithwick2011}) and the other, to more quiescent ones (e.g., disk migration and in-situ formation, \citealt{Goldreich1980,Lin1996,Ward1997,Batgyin2016}). Orbital eccentricity provides insight into a planet’s dynamical history; however, tidal circularization may erase that evidence. Warm Jupiters ($M$ = 0.3-13$\mj$), with a larger semi-major axis ($a$ = 0.1-1 AU), are less influenced by tidal dissipation, preserving any evolutionary eccentricity excitation \citep{Dawson2018}. This creates a unique observational window to study the dynamical histories of short-period gas giants. 

When comparing observed warm Jupiter eccentricities to those expected from the two major formation pathways, many cannot be easily described by either mechanism; their eccentricities falling in between the two regions of expectation. These warm Jupiters are too circular for high-eccentricity migration (e > 0.9) while simultaneously too eccentric for either disk migration or in-situ formation \citep[e < 0.2;][]{2004ApJ...614..497G}. This phenomenon of being too circular yet too eccentric requires either a new mechanism or closer inspection of currently-known mechanisms to explain these warm Jupiters' moderate eccentricities. 

\begin{figure*}
        \centering
        \includegraphics[width=\textwidth]{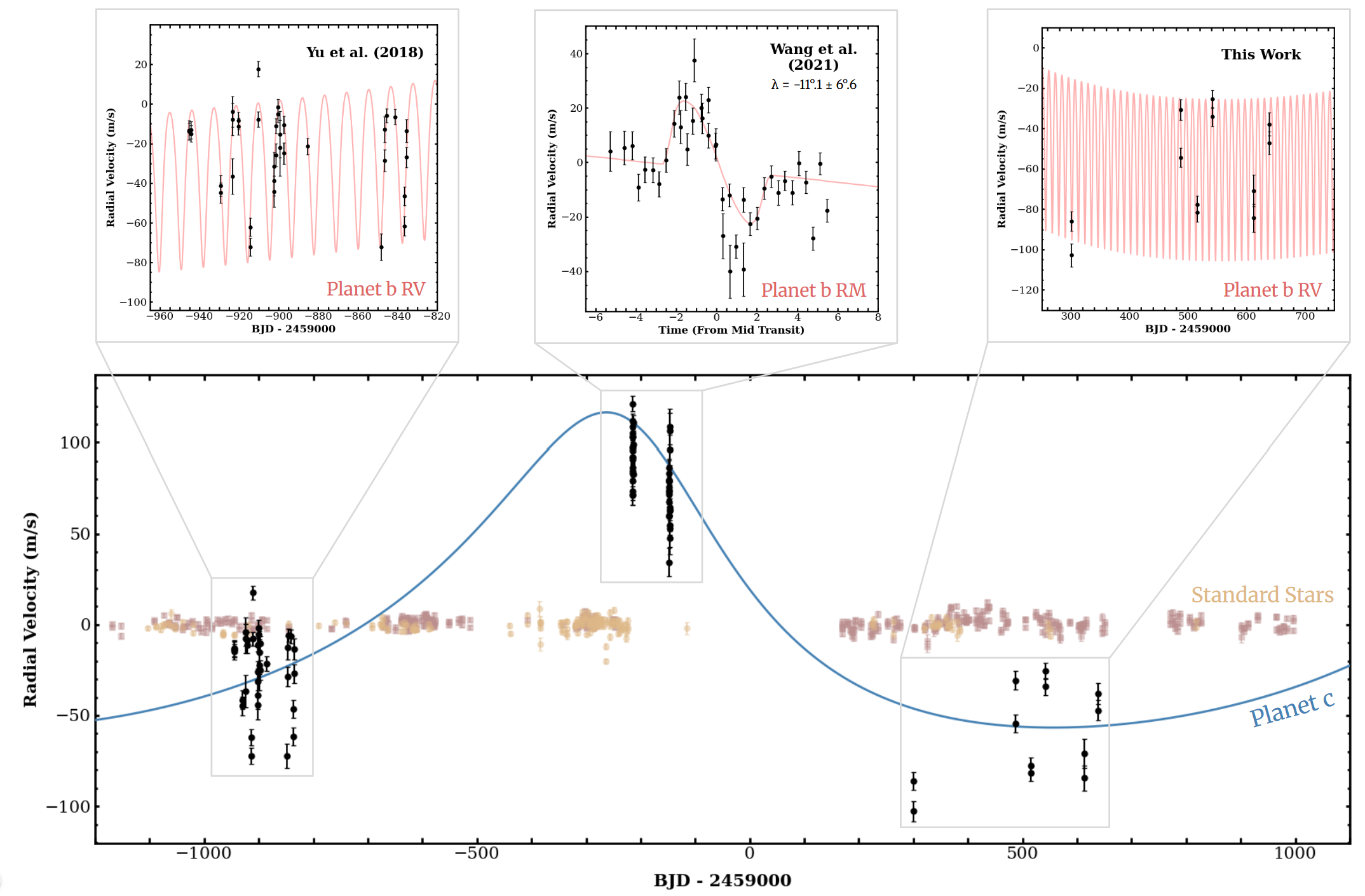}
        \caption{APF RV data demonstrating a lack of instrumental systematics and suggesting the presence of a cold Jupiter in the K2-232 system. Top left panel: RV measurements for K2-232 with the best-fit curve for planet b overplot in red from the \citealt{Yu2018} discovery paper. Top middle panel: RM effect RV measurements versus time from mid-transit for K2-232 with the best-fit curve overplot in red from \citealt{Wang2021}. Top right panel: RV measurements obtained in this work for K2-232 with the best-fit curve overplot in red. Bottom panel: All APF RV measurements for K2-232; the blue trend line is indicative of a long-period giant companion responsible for the large RV offsets between past RV campaigns for this system. The mauve and tan data points show standard star measurements used to confirm the lack of instrumentation error (HD 9407 - mauve square points (RMS$_{RV} = 3.13$ m/s), HD 199305 - tan circle points (RMS$_{RV} = 3.18$ m/s)).
}
        \label{fig:btfr}
\end{figure*}

High-eccentricity migration could describe warm Jupiters with intermediate eccentricities if an outer planetary companion ($M$ = 0.3-13$\mj$, $a > 1$ AU) is also present. Postulated in \citealt{2014ApJ...781L...5D}, an outer gas giant can drive eccentricity oscillations of its inner companion through secular planet-planet interactions over timescales of $10^4-10^5$ years \citep{Petrovich2016}. This suggests that some moderately eccentric warm Jupiters are on a high-eccentricity migration path to becoming hot Jupiters and are simply being observed in a phase of lower eccentricity. 

Although this is a potentially notable explanation for the formation of more eccentric warm Jupiters, observational constraints on this model are limited by the small sample size of known warm Jupiter-cold Jupiter pairs (e.g.,  Kepler-432, \citealt{Quinn2015} and Kepler-419, \citealt{dawson2014}).

Most warm Jupiter data comes from bright stars in wide-field surveys or faint stars in limited coverage ones; both complicating follow-up observations. To find warm Jupiters, we must observe as many stars as possible due to their low occurrence rate ($\sim1 \%$, \citealt{Fulton2021}) and their diminished transit probability as compared to closer-in hot Jupiters \citep[P$_{tra} \propto a^{-1}$,][]{Winn2010b}. Wide-field surveys from ground-based telescopes such as SuperWASP \citep{Pollacco2006}, KELT \citep{KELT}, and HAT \citep{Bakos2004} are an effective avenue to discovering hot Jupiters, however, they cannot perform continuous observations due to weather and day-night cycles, making it difficult to build the long-term baselines needed to detect warm Jupiters.

Considering warm Jupiters are intrinsically rare \citep{Jones2003,Dawson2018} and it is not likely to find numerous bright stars ($K_P$ < 13) in a given area of the sky, warm Jupiters orbiting bright stars within \kepler\,'s small field of view \citep{2009IAUS..253..289B} were rarely observed. Instead, most of \kepler\,'s warm Jupiters were discovered around dimmer stars (14 < $K_P$ < 16) which made them difficult to follow up. 

\begin{deluxetable}{lcc}
\tablecaption{Astrometry and Photometry for K2-232\label{tab:toi2005}}
\tablewidth{0.5\columnwidth}
\tablehead{
\colhead{{Parameter}} & \colhead{{Value}} & \colhead{{Source}}
}
\startdata
\multicolumn{3}{l}{\textbf{Catalog Information}} \\
TIC ID & TIC 68577662 & TOI Catalog \\
TOI ID & TOI-4606 & TOI Catalog \\
Gaia DR3 ID & 3406687485600728192 & Gaia DR3 \\
2MASS ID & J04550395+1839164 & 2MASS \\
\multicolumn{3}{l}{\textbf{Coordinates and Proper Motion}} \\
Right Ascension & 04:55:04.03 & Gaia DR3 \\
% Right Ascension (hh:mm:ss) & 04:55:04.03 & Gaia DR3 \\
% Declination (dd:mm:ss) & +18:39:15.56 & Gaia DR3 \\
Declination & +18:39:15.56 & Gaia DR3 \\
Parallax (mas) & 7.6214$\pm$0.0443369 & Gaia DR3 \\
$\mu_{\mathrm{R.A}}$ (mas yr$^{-1}$) & 62.0635$\pm$0.0766353 & Gaia DR3 \\
$\mu_{\mathrm{Dec.}}$ (mas yr$^{-1}$) & $-48.2449\pm0.0507786$ & Gaia DR3 \\[3pt]
\multicolumn{3}{l}{\textbf{Magnitudes}} \\
$G$ (mag) & $9.731985 \pm 0.002759$ & Gaia DR3 \\
$B_P$ (mag) & $10.036352 \pm 0.002804$ & Gaia DR3 \\
$R_P$ (mag) & $9.259789 \pm 0.003784$ & Gaia DR3 \\
$J$ (mag) & $8.739 \pm 0.0300$ & 2MASS \\
$H$ (mag) & $8.480 \pm 0.0180$ & 2MASS \\
$K$ (mag) & $8.434 \pm 0.0170$ & 2MASS \\
$WISE_{3.4\mu m}$ (mag) & $8.380 \pm 0.0240$ & WISE \\
$WISE_{4.6\mu m}$ (mag) & $8.419 \pm 0.0190$ & WISE \\
$WISE_{12\mu m}$ (mag) & $8.391 \pm 0.0270$ & WISE \\
% $WISE_{22\mu}$ (mag) & $8.605 \pm $No error found & WISE \\
\enddata

\tablenotetext{}{Note: TESS TOI Primary Mission Catalog: \citet{Guerrero2021}, GAIA DR3: \citet{Gaiadr32022arXiv220800211} , 2MASS: \citet{Cutri2003}, WISE: \citet{WISE} }
\end{deluxetable}

The landscape for the discovery of warm Jupiter systems completely changed with \ktwo\, and \tess\, \citep{Howell2014, Ricker2015}. With both being space-based, wide-field surveys, they were able to find numerous warm Jupiters. Of those discovered, some were even found around bright stars \citep{2021ApJS..255....6D, 2021AJ....161..194R, Brahm2023, Andres2020}. These bright stars are ideal for ground-based radial velocity (RV) follow up as producing high resolution spectra is necessary to search for those potentially dynamically significant long-period, gas giant companions. RVs are also sensitive to the signal of a planet across its entire orbit, so the weather and day-night cycles that prevent warm Jupiter detection with ground-based transit surveys do not restrict ground-based RV surveys in the same way. 

Still, there remain logistical barriers for warm Jupiter follow up observations. While the exoplanet census is expanding on a near-daily basis, the number of telescopes and instruments available to confirm and characterize these planets is not. So, many systems are only given enough telescope time for confirmation. Thus, a majority of systems are seldom followed up in search for outer companions, limiting the discovery of cold Jupiters, which require long baseline, RV data as their wide orbits generally mean they will not transit their host star as observed from Earth. Even when these systems are the subject of sustained RV follow-up campaigns, confirmation of additional, non-transiting, long-period planets is challenging. Instrumental drifts or zero-point offsets can mimic or obscure the presence of long period RV trends, even when a single RV spectrograph is used for the entire campaign. For systems where multiple RV spectrographs are used but the data sets do not overlap in time, an additional RV offset term must be fit for each set of RVs which makes the detection of long-period planets significantly harder. Together, these issues make it difficult to search for cold Jupiters in known warm Jupiter systems. Yet, we are continuing to explore this complicated regime with the discovery of K2-232c.

The history of K2-232 begins with the discovery of its warm Jupiter. Both \citealt{Brahm2018} and \citealt{Yu2018} simultaneously discovered K2-232b, a transiting warm Jupiter ($P$ = 11 days) orbiting a late F-type star ($V$ = 9.9, $v$sin\textit{i} = 4.16 km/s). Given the stellar brightness, the star's projected rotational velocity, and the presence of a transit ($T_{1-4}$ = 5.03 hours), the system provided an excellent opportunity for Rossiter-McLaughlin (RM) effect observations \citep{rossiter1924detection, mclaughlin1924some} which determine a system's spin-orbit angle. \citealt{Wang2021} performed such measurements to find the normal to the planet's orbital plane was well-aligned with the spin axis of its host star. However, the RV data taken in \citealt{Wang2021} and \citealt{Yu2018} were significantly offset from each other even though both used the same instrument and there was no significant instrument or pipeline interventions between the two data sets. This indicated the possible existence of another planet in the system with a larger semi-major axis (See \autoref{fig:btfr}).

In this paper, we present additional RV data for K2-232 that reveals the presence of an eccentric cold Jupiter (K2-232c; $e=0.35$) in this known warm Jupiter system that may help us understand the complicated formation path of warm Jupiters. With an eccentricity of $0.245 ^{+0.023}_{-0.024}$, if isolated, the inner warm Jupiter's formation would be best explained by either disk migration or in-situ formation. However, with the companion we are introducing in this work, the high-eccentricity migration pathway can also explain the presence of the inner planet via low-eccentricity phasing.

\begin{figure}
        \centering
        \includegraphics[width=\columnwidth]{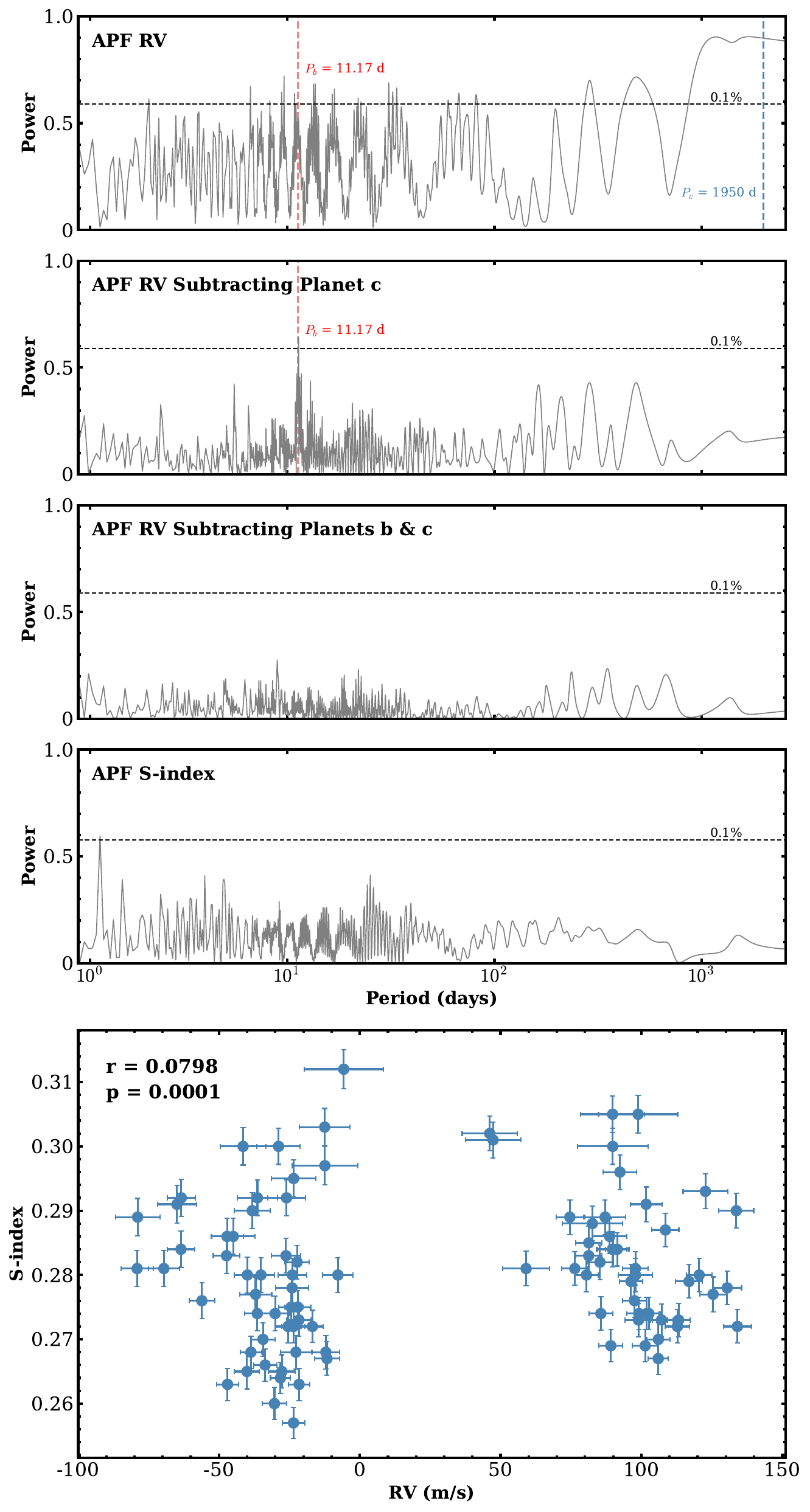}
        \caption{Top panel: Periodogram of all APF RV measurements. The blue vertical dashed line indicates where the period of planet c dominates and the red vertical dashed line indicates where the period of planet b dominates. Second panel: RV periodogram after subtracting planet c's trend. Only planet b's period remains dominant. Third panel: RV periodogram after subtracting both planet b and c's trend. No period meets the power criteria necessary for another object in the system. Fourth panel: Periodogram of all APF S-index measurements. There is no significant power found at either planet b or c's period indicating a lack of stellar magnetic activity on either timescale. The black horizontal dashed line in each of the above panels indicates a False Alarm Probability of 0.1\%. Bottom panel: Scatter plot of APF S-index measurements versus their corresponding residual APF RV measurements. Both the Pearson correlation coefficient ($r$) and \textit{p}-value ($p$) are shown in the top left of the panel. Together, these values indicate a statistically significant non-correlation in the data.}
        \label{fig:sindex}
\end{figure}

In the following sections, we describe both our photometric and spectroscopic observations of K2-232 (Section \ref{sec:obs}), the fit we performed to discover and analyze the outer companion (Section \ref{sec:glob}), and what those results suggest for warm Jupiter formation (Section \ref{sec:disc}).

\section{Observations}
\label{sec:obs}

\begin{deluxetable}{ccc}
\tablecaption{Summary of all observation details carried out for the K2-232 system\label{tab:obsdetail}}
\tablewidth{0.5\columnwidth}
\tablehead{
\multicolumn{3}{c}{\textbf{Photometric Observations}} \\
\colhead{{\textbf{Observatory}}} & \colhead{{\textbf{Campaigns/Sectors}}} & \colhead{{\textbf{Number of Transits}}}
}
\startdata
K2 & Campaign 13 & 7 \\
TESS & Sectors 43, 44, 71 & 2 \\
\hline
\multicolumn{3}{c}{\textbf{Spectroscopic Observations}} \\
\textbf{Instrument} & \textbf{Date} & \textbf{Number of Spectra} \\
\hline
CORALIE/Euler 1.2 m & 2017 October 31 - 2017 November 2 & 3 \\
% Right Ascension (hh:mm:ss) & 04:55:04.03 & Gaia DR3 \\
% Declination (dd:mm:ss) & +18:39:15.56 & Gaia DR3 \\
FEROS/MPG 2.2 m & 2017 October 3 - 2018 January 28 & 18 \\
HARPS/ESO 3.6 m & 2017 November 1 - 2017 November 8 & 8 \\
Levy/APF 2.4 m & 2017 October 28 - 2017 December 27 & 19 \\
& 2019 October 29 - 2020 January 4 & 40 \\
& 2021 March 27 - 2022 February 28 & 12 \\
\enddata
\end{deluxetable}

In this project, the majority of our data were sourced from archival photometric and spectroscopic measurements. Here, we discuss where that information originated and the additional data we gathered independently. 

\subsection{Photometric Observations}

We sourced photometric data from the \ktwo\, \citep{Howell2014} and \tess\, \citep{Ricker2015} missions. Observations of K2-232 were conducted by the \ktwo\, mission during Campaign 13 (2017 March 12 - 2017 May 19) and by the \tess\, mission in Sectors 43, 44, and 71 (2021 September 16 - 2021 November 5 and 2023 October 20 - 2023 November 10). The \ktwo\, and \tess\, data were retrieved from the Mikulski Archive for Space Telescopes (MAST\footnote[1]{https://archive.stsci.edu/}) using \texttt{lightkurve} \citep{2018ascl.soft12013L}. The specific observations analyzed can be accessed via \dataset[doi:10.17909/tc6v-yd53]{https://doi.org/10.17909/tc6v-yd53}.

The \ktwo\, mission data were processed using the \texttt{EVEREST} algorithm \citep{Luger2016, Luger2018}, which applies a variant of pixel-level decorrelation \citep{Deming2015} to correct for systematics caused by spacecraft pointing variations, along with a Gaussian process to model astrophysical variability. Regarding the \tess\, mission light curves, we adopted the 2-minute cadence Pre-search Data Conditioning Simple Aperture Photometry \cite[PDCSAP;][]{Smith2012PDCSAP,Stumpe2012, Stumpe2014} light curve, developed by the Science Processing Operation Center (SPOC) team \citep{Jenkins2016}. This involves correcting for both long-term and systematic trends. First, masking the transits out of the data, we then detrended the \tess\, light curves using Gaussian processes with a Matern 3/2 kernel, implemented in \texttt{wotan} \citep{Hippke2019}. We omitted any data points with critical quality concerns.

% All the photometric transit data we gathered for the K2-232 system came from either its discovery telescope, \textit{K2} \citep{Howell2014}, or from recent sectors of the Transiting Exoplanet Survey Satellite (\textit{TESS}). Using lightkurve (\citealt{2018ascl.soft12013L}), we downloaded this transit data from the Mikulski Archive for Space Telescopes (MAST) \footnote{https://archive.stsci.edu/} to use in our global fitting. 

% From \textit{K2}, we utilized data from Campaign 13 released in 2017 via the EVEREST pipeline (\citealt{2016AJ....152..100L}) and from \textit{TESS}, we compiled data from Sectors 43 and 44, both released in 2021 through the SPOC pipeline (\citealt{Jenkins2016}). 

\subsection{Spectroscopic Follow-up}
\label{sec:specobs}

% In \citealt{Brahm2018}, the authors collected data from three spectrographs, each mounted on separate telescopes at the ESO La Silla Observatory. They took 3 spectra with the CORALIE spectrograph \citep{2001Msngr.105....1Q} on the 1.2-m Euler/Swiss telescope from 2017 October 31 to 2017 November 2. \todo{CORALIE explanation here} 18 more spectra were taken with the FEROS spectrograph \citep{1999Msngr..95....8K} \textendash{} mounted on the MPG 2.2-m telescope \textendash{} between 2017 October 3 and 2018 January 28. \todo{FEROS explanation here} Then from 2017 November 1 to 2017 November 8, the authors took an additional 8 spectra with the HARPS spectrograph \citep{Mayor2003} mounted to the ESO 3.6-m telescope. \todo{HARPS explanation here} 

The discovery paper, \citealt{Brahm2018}, presented 29 RV measurements obtained with three spectrographs: CORALIE/Euler 1.2 m  \citep{2001Msngr.105....1Q}, FEROS/MPG 2.2 m \citep{1999Msngr..95....8K}, and HARPS/ESO 3.6 m \citep{Mayor2003}. The CORALIE RV observations were conducted in simultaneous calibration mode \citep{Baranne1996}, with a Fabry Perot etalon illuminating the secondary fiber to trace instrumental velocity drifts. Both FEROS and HARPS observations were also carried out using the simultaneous calibration mode, but with a Thorium Argon lamp illuminating the secondary fiber. For the above three datasets, RVs were extracted using the cross correlation technique as implemented in the CERES automated pipeline \citep{Jordan2014,Brahm2017CERES}. 

Concurrently, the authors from the simultaneous discovery paper \citep{Yu2018} collected 19 RVs for the system with Levy/APF 2.4 m, located at Lick Observatory \citep{Vogt2014}. The APF is a fully automated facility, which carries out nightly observations using a dynamic scheduler based on \citealt{Burt2014} and \citealt{Fulton2015}. It measures RVs via a cell of gaseous I$_2$ placed in the converging beam of the telescope. This cell imprints the 5000-6200\AA\ region of incoming stellar spectrum with a dense forest of I$_2$ lines that act as a wavelength calibrator and provide a proxy for the point spread function (PSF) of the spectrometer. To ensure a constant I$_2$ column density over multiple decades, the cell is held at a constant temperature of 50.0 $\pm$ 0.1 $^{\circ}$C. The APF has a typical spectral resolution of 90,000 and produces spectra from 3700 - 7700\AA, though only the 5000-6200\AA\ spectral region is used for measuring RVs. 

\citealt{Wang2021} targeted the K2-232 system further as it made a quality candidate for observing the RM effect \citep{rossiter1924detection, mclaughlin1924some} with both transit and RV data available. Again using the APF, the authors took RV data on two nights covering two separate transit events.

In this paper, we report 12 new APF RV measurements. We reprocess all of the existing APF observations using the data reduction pipeline maintained at the Carnegie Earth and Planets Laboratory to ensure homogeneous treatment of the data. Raw APF data is reduced to 1D wavelength-calibrated spectra with a custom built raw reduction package. Once the iodine region of the spectrum has been extracted, it is split into 2\AA\ chunks. Each chunk is analyzed using the spectral synthesis technique described in \citealt{Butler1996}, which deconvolves the stellar spectrum from the I$_2$ absorption lines and produces an independent measure of the wavelength, instrument PSF, and Doppler shift. The final Doppler velocity from a given observation is the weighted mean of the velocities of all the individual chunks ($\sim$700 for the APF). The final internal uncertainty of each velocity is the standard deviation about the weighted mean for all 700 chunk velocities. See \autoref{tab:obsdetail} for all specific observation details.

\providecommand{\tjdtdb}{\ensuremath{\rm {TJD_{TDB}}}}

\providecommand{\bjdtdb}{\ensuremath{\rm {BJD_{TDB}}}}
\providecommand{\feh}{\ensuremath{\left[{\rm Fe}/{\rm H}\right]}}
\providecommand{\teff}{\ensuremath{T_{\rm eff}}}
\providecommand{\teq}{\ensuremath{T_{\rm eq}}}
\providecommand{\ecosw}{\ensuremath{e\cos{\omega_*}}}
\providecommand{\esinw}{\ensuremath{e\sin{\omega_*}}}
\providecommand{\msun}{\ensuremath{\,M_\Sun}}
\providecommand{\rsun}{\ensuremath{\,R_\Sun}}
\providecommand{\lsun}{\ensuremath{\,L_\Sun}}
\providecommand{\mj}{\ensuremath{\,M_{\rm J}}}
\providecommand{\rj}{\ensuremath{\,R_{\rm J}}}
\providecommand{\me}{\ensuremath{\,M_{\rm E}}}
\providecommand{\re}{\ensuremath{\,R_{\rm E}}}
\providecommand{\fave}{\langle F \rangle}
\providecommand{\fluxcgs}{10$^9$ erg s$^{-1}$ cm$^{-2}$}

%\usepackage{apjfonts}
%\begin{document}

\startlongtable
\begin{deluxetable*}{lccccccc}
\tablecaption{
Median Parameter Values and 68\% Confidence Intervals}
\tabletypesize{\scriptsize}    
\tablehead{\colhead{~~~Parameter} & \colhead{Units} & \multicolumn{6}{c}{Values}}
%\tablehead{\colhead{~~~Parameter} & \colhead{Description} & %\multicolumn{6}{c}{Values}}
\startdata
\multicolumn{2}{l}{Priors:}&\smallskip\\
~~~~$\varpi$\dotfill &Parallax (mas)\dotfill &  \multicolumn{2}{c}{$\mathcal{G}(7.73883, 0.01972)$}  \\
~~~~$[{\rm Fe/H}]$\dotfill &Metallicity (dex)\dotfill & \multicolumn{2}{c}{$\mathcal{G}(0.10, 0.04)$}\\
~~~~$A_V$\dotfill &V-band extinction (mag)\dotfill &\multicolumn{2}{c}{$\mathcal{U}(0, 1.93)$}\\
\multicolumn{2}{l}{Stellar Parameters:}&\medskip\\
~~~~$M_*$\dotfill &Mass (\msun)\dotfill &\multicolumn{2}{c}{$1.121^{+0.063}_{-0.065}$}\\
~~~~$R_*$\dotfill &Radius (\rsun)\dotfill &\multicolumn{2}{c}{$1.210^{+0.036}_{-0.035}$}\\
%~~~~$R_{*,SED}$\dotfill &Radius$^{1}$ (\rsun)\dotfill &$1.178^{+0.011}_{-0.010}$\\
%~~~~$L_*$\dotfill &Luminosity (\lsun)\dotfill &$1.84\pm0.13$\\
%~~~~$F_{Bol}$\dotfill &Bolometric Flux (cgs)\dotfill &$3.52\pm0.26 \times 10^{-9}$\\
%~~~~$\rho_*$\dotfill &Density (cgs)\dotfill &$0.891^{+0.092}_{-0.086}$\\
~~~~$\log{g}$\dotfill &Surface gravity (cgs)\dotfill &\multicolumn{2}{c}{$4.322^{+0.033}_{-0.035}$}\\
~~~~$T_{\rm eff}$\dotfill &Effective Temperature (K)\dotfill &\multicolumn{2}{c}{$6100\pm100$}\\
%~~~~$T_{\rm eff,SED}$\dotfill &Effective temperature$^{1}$ (K)\dotfill &$6190^{+130}_{-140}$\\
~~~~$[{\rm Fe/H}]$\dotfill &Metallicity (dex)\dotfill &\multicolumn{2}{c}{$0.106\pm0.039$}\\
%~~~~$[{\rm Fe/H}]_{0}$\dotfill &Initial Metallicity$^{2}$ \dotfill &$0.137^{+0.042}_{-0.044}$\\
~~~~$Age$\dotfill &Age (Gyr)\dotfill &\multicolumn{2}{c}{$3.9^{+2.4}_{-1.9}$}\\
%~~~~$EEP$\dotfill &Equal Evolutionary Phase$^{3}$ \dotfill &$378^{+30}_{-35}$\\
~~~~$A_V$\dotfill &V-band extinction (mag)\dotfill &\multicolumn{2}{c}{$0.237^{+0.081}_{-0.091}$}\\
%~~~~$\sigma_{SED}$\dotfill &SED photometry error scaling \dotfill &$0.57^{+0.24}_{-0.15}$\\
~~~~$\varpi$\dotfill &Parallax (mas)\dotfill &\multicolumn{2}{c}{$7.739\pm0.020$}\\
~~~~$d$\dotfill &Distance (pc)\dotfill &\multicolumn{2}{c}{$129.22\pm0.33$}\\
\multicolumn{2}{l}{Planetary Parameters:}&b&c\medskip\\
~~~~$P$\dotfill &Period (days)\dotfill &$11.1684377\pm0.0000010$&$1950^{+140}_{-120}$\\
~~~~$R_P$\dotfill &Radius (\rj)\dotfill &$1.037^{+0.031}_{-0.030}$&$-$\\
~~~~$M_P$\dotfill &Mass (\mj)\dotfill &$0.427^{+0.039}_{-0.036}$&$-$\\
~~~~$M_P\sin i$\dotfill &Minimum mass (\mj)\dotfill &$0.427^{+0.039}_{-0.036}$&$5.31^{+0.48}_{-0.45}$\\
~~~~$M_P/M_*$\dotfill &Mass ratio \dotfill &$0.000364^{+0.000032}_{-0.000030}$&$0.00453^{+0.00039}_{-0.00036}$\\
~~~~$T_C$\dotfill &Observed Time of Conjunction (\bjdtdb)\dotfill &$2458160.404632\pm0.000048$&$-$\\
% ~~~~$T_C$\dotfill &Model Time of Conjunction$^{4,5}$ (\tjdtdb)\dotfill &$2458160.404076^{+0.000054}_{-0.000055}$&$-$\\
%~~~~$T_T$\dotfill &Model time of min proj sep$^{5,6,7}$ (\tjdtdb)\dotfill &$2457981.709080\pm0.000051$&--\\
%~~~~$T_0$\dotfill &Obs time of min proj sep$^{6,8,9}$ (\bjdtdb)\dotfill &$2457981.709635^{+0.000044}_{-0.000043}$&$2.458982^{+0.000030}_{-0.000028} \times 10^{6}$\\
~~~~$a$\dotfill &Semi-major Axis (AU)\dotfill &$0.1016^{+0.0019}_{-0.0020}$&$3.17^{+0.16}_{-0.14}$\\
~~~~$i$\dotfill &Inclination (Degrees)\dotfill &$89.83^{+0.12}_{-0.18}$&$88.65^{+0.95}_{-1.5}$\\
~~~~$e$\dotfill &Eccentricity \dotfill &$0.245^{+0.023}_{-0.024}$&$0.352^{+0.095}_{-0.076}$\\
~~~~$\omega_*$\dotfill &Argument of Periastron (Degrees)\dotfill &$-178.2^{+7.6}_{-7.7}$&$17.9^{+9.7}_{-11}$\\
%~~~~$T_{\rm eq}$\dotfill &Equilibrium temp$^{10}$ (K)\dotfill &$1016\pm17$&$181.8\pm4.9$\\
%~~~~$\tau_{\rm circ}$\dotfill &Tidal circ timescale (Gyr)\dotfill &$20.1^{+4.4}_{-3.9}$&$4.1^{+3.9}_{-2.4} \times 10^{11}$\\
~~~~$K$\dotfill &RV Semi-amplitude (m/s)\dotfill &$37.2^{+3.3}_{-3.0}$&$85.6^{+5.1}_{-3.8}$\\
~~~~$R_P/R_*$\dotfill &Radius of planet in stellar radii \dotfill &$0.08799^{+0.00014}_{-0.00013}$&$-$\\
~~~~$a/R_*$\dotfill &Semi-major axis in stellar radii \dotfill &$18.04\pm0.60$&$564^{+31}_{-29}$\\
%~~~~$\delta$\dotfill &$\left(R_P/R_*\right)^2$ \dotfill &$0.007743^{+0.000024}_{-0.000022}$&$0.0092^{+0.0015}_{-0.0014}$\\
%~~~~$\delta_{\rm Kepler}$\dotfill &Transit depth in Kepler (frac)\dotfill &$0.009713^{+0.000092}_{-0.000091}$&$0.00$\\
%~~~~$\delta_{\rm TESS}$\dotfill &Transit depth in TESS (frac)\dotfill &$0.00874^{+0.00013}_{-0.00012}$&$0.00$\\
%~~~~$\tau$\dotfill &In/egress transit duration (days)\dotfill &$0.016964^{+0.00016}_{-0.000046}$&$0.00$\\
~~~~$T_{14}$\dotfill &Total transit duration (days)\dotfill &$0.20911^{+0.00027}_{-0.00026}$&$-$\\
%~~~~$T_{FWHM}$\dotfill &FWHM transit duration (days)\dotfill &$0.19210^{+0.00026}_{-0.00025}$&$0.00$\\
~~~~$b$\dotfill &Transit impact parameter \dotfill &$0.052^{+0.055}_{-0.036}$&$10.2^{+11}_{-7.2}$\\
~~~~$T_P$\dotfill &Time of Periastron (\tjdtdb)\dotfill &$2458151.22^{+0.25}_{-0.23}$&$2458786^{+40}_{-63}$\\
%~~~~$T_A$\dotfill &Time of asc node (\tjdtdb)\dotfill &$2458156.72^{+0.18}_{-0.17}$&$2.46068^{+0.00015}_{-0.00011} \times 10^{6}$\\
%~~~~$T_D$\dotfill &Time of desc node (\tjdtdb)\dotfill &$2458162.36^{+0.12}_{-0.11}$&$2.459567^{+0.000089}_{-0.000097} \times 10^{6}$\\
%~~~~$V_c/V_e$\dotfill &Scaled velocity \dotfill &$0.977\pm0.033$&$0.846^{+0.079}_{-0.095}$\\
% ~~~~$e\cos{\omega_*}$\dotfill & \dotfill &$-0.243^{+0.024}_{-0.023}$&$0.333^{+0.074}_{-0.067}$\\
% ~~~~$e\sin{\omega_*}$\dotfill & \dotfill &$-0.008^{+0.033}_{-0.032}$&$0.107^{+0.088}_{-0.073}$\\
%~~~~$d/R_*$\dotfill &Separation at mid transit \dotfill &$17.1^{+1.2}_{-1.1}$&$446^{+54}_{-65}$\\
%~~~~$P_T$\dotfill &A priori non-grazing transit prob \dotfill &$0.0533^{+0.0038}_{-0.0035}$&$0.00202^{+0.00035}_{-0.00022}$\\
%~~~~$P_{T,G}$\dotfill &A priori transit prob \dotfill &$0.0636^{+0.0046}_{-0.0041}$&$0.00245^{+0.00042}_{-0.00026}$\\
%~~~~$P_S$\dotfill &A priori non-grazing eclipse prob \dotfill &$0.05415^{+0.0010}_{-0.00096}$&$0.00164^{+0.00012}_{-0.00010}$\\
%~~~~$P_{S,G}$\dotfill &A priori eclipse prob \dotfill &$0.0646^{+0.0012}_{-0.0011}$&$0.00199^{+0.00013}_{-0.00012}$\\
\multicolumn{2}{l}{Wavelength Parameters:}&Kepler&TESS\smallskip\\
~~~~$u_{1}$\dotfill &Linear limb-darkening coeff \dotfill &$0.408\pm0.013$&$0.230\pm0.026$\\
~~~~$u_{2}$\dotfill &Quadratic limb-darkening coeff \dotfill &$0.207\pm0.029$&$0.291\pm0.040$\\
\multicolumn{2}{l}{Telescope Parameters:}&APF&CORALIE\smallskip\\
~~~~$\gamma_{\rm rel}$\dotfill &Relative RV Offset (m/s)\dotfill &$-20.6^{+7.4}_{-7.2}$&$22371\pm33$\\
~~~~$\sigma_J$\dotfill &RV Jitter (m/s)\dotfill &$12.1^{+1.3}_{-1.1}$&$85^{+220}_{-52}$\\
~~~~$\sigma_J^2$\dotfill &RV Jitter Variance \dotfill &$147^{+33}_{-26}$&$7300^{+84000}_{-6200}$\\
\multicolumn{2}{l}{}&FEROS&HARPS\smallskip\\
~~~~$\gamma_{\rm rel}$\dotfill &Relative RV Offset (m/s)\dotfill &$22423.6^{+8.1}_{-7.7}$&$22424.7^{+8.7}_{-8.4}$\\
~~~~$\sigma_J$\dotfill &RV Jitter (m/s)\dotfill &$6.5\pm3.7$&$2.1^{+3.6}_{-2.1}$\\
~~~~$\sigma_J^2$\dotfill &RV Jitter Variance \dotfill &$42^{+61}_{-35}$&$4^{+28}_{-10}$\\
%\smallskip\\\multicolumn{2}{l}{Transit Parameters:}&k2232 UT 2024-05-18 (Kepler)&k2-232 UT 2024-05-18 (TESS)\smallskip\\
%~~~~$\sigma^{2}$\dotfill &Added Variance \dotfill &$-1.27098^{+0.00028}_{-0.00023} \times 10^{-6}$&$3.9^{+2.1}_{-2.0} \times 10^{-8}$\\
%~~~~$F_0$\dotfill &Baseline flux \dotfill &$1.0000028\pm0.0000054$&$1.000093\pm0.000021$\\
\enddata
\label{tab:K2-232_exofast.}
% \tablenotetext{}{See Table 3 in \citet{Eastman:2019} for a detailed description of all parameters}
% \tablenotetext{1}{This value ignores the systematic error and is for reference only}
% \tablenotetext{2}{The metallicity of the star at birth}
% \tablenotetext{3}{Corresponds to static points in a star's evolutionary history. See \S2 in \citet{Dotter:2016}}
% \tablenotetext{4}{Time of conjunction is commonly reported as the ``transit time''}
\tablenotetext{}{Note-\tjdtdb\, is the target's barycentric frame and corrects for light travel time.}

% \tablenotetext{6}{Time of minimum projected separation is a more correct ``transit time''}
% \tablenotetext{7}{Use this to model TTVs, e}
% \tablenotetext{8}{At the epoch that minimizes the covariance between $T_C$ and Period}
% \tablenotetext{9}{Use this to predict future transit times}
% \tablenotetext{10}{Assumes no albedo and perfect redistribution}
\end{deluxetable*}
% \bibliographystyle{apj}
% \bibliography{References}
%\end{document}

\section{Fitting and Parametrization} 
\label{sec:glob}

\subsection{Discovery \& Confirmation K2-232c}

The discovery of K2-232c arose from investigation into the alignment of the inner warm Jupiter companion, K2-232b. \citealt{Wang2021} took RV measurements during two separate transit events (2019 October 29 and 2020 January 4) to find the planet’s projected spin–orbit angle via the RM effect. Through these two measurements, planet b's orbit was found to be well-aligned with the spin axis of the host star ($\lambda = -11\degrees.1\pm6\degrees.6$, top middle panel in \autoref{fig:btfr}). However, the RV data taken in \citealt{Wang2021} and the out-of-transit RVs taken in the discovery paper (\citealt{Yu2018}; top left panel in \autoref{fig:btfr}) were offset from one another. Although both groups used the same instrument on the same telescope, their data had an RV offset of $\sim$100 $\mathrm{m/s}$. To check for possible instrumental or data reduction pipeline systematics, we overlaid measurements from two RV standard stars observed over the same time span (HD 9407 shown by square mauve points and HD 199305 shown by circle tan points in the bottom panel of \autoref{fig:btfr}). Finding no offsets within the standard stars' RV time series (RMS$_{RV} = 3.13$ m/s and RMS$_{RV} = 3.18$ m/s respectively), we collected more RV data with the APF (see \autoref{sec:specobs}). We plotted our new RVs in \autoref{fig:btfr} (top right panel), along with all other APF RV data (bottom panel).

Using our newly obtained RV measurements as well as all those already present in the literature, we analyzed the possible planetary signals in the RVs using the Lomb–Scargle \citep{Lomb, Scargle} periodogram (first four panels of \autoref{fig:sindex}). After finding and removing a periodic signal at 11.17 days (second panel in \autoref{fig:sindex}) and 1950 days (first panel in \autoref{fig:sindex}), there was no significant power that met the criteria necessary (>0.1\% False Alarm Probability; \citealt{Zechmeister2009}) to indicate the presence of another companion in the system (third panel in \autoref{fig:sindex}). 

To check whether the velocities we measured were associated with the star's magnetic cycle \citep{Dumusque2011}, we first created another periodogram using only S-index measurements from our APF observations (fourth panel in \autoref{fig:sindex}). With no notable peaks at either planet b or c’s possible period, it is more likely that the long term signal is due to an outer planet in the system. However, to strengthen this claim, we also compared our S-index values to our residual RVs and subsequently plotted this in the bottom panel of \autoref{fig:sindex}. Using the Pearson correlation coefficient to quantify the relationship of our variables ($r > 0.30$ for a moderate linear relationship) as well as a \textit{p}-value to test the relationship's statistical significance ($p<0.05$), we found no correlation between our two sets of values ($r = 0.0798$) which was statistically validated ($p = 0.0001$). Having ruled out instrumental drifts, data reduction pipeline systematics, and a stellar magnetic cycle as possible sources for this additional, long period signal, we are left with the conclusion that it is instead due to an outer planetary companion, hereafter identified as K2-232 c.

% EXOFASTv2 uses a differential evolution Markov Chain Monte Carlo (DE-MCMC) algorithm (\citealt{TerBraak}) combined with the AMOEBA symplectic solver to simultaneously fit transit light curves with RVs. 

% For this fit, we use photometry from both TESS and K2 with RVs from APF-Levy, ESO-CORALIE, ESO-FEROS, and ESO-HARPS.

\subsection{EXOFASTv2 Global Fitting}
\subsubsection{Acquiring Stellar Parameters}
\label{subsec:star}

% To derive the stellar parameters of K2-232, we fit the spectral energy distribution (SED) using MESA Isochrones and Stellar Tracks (MIST; \citealt{Choi2016mist, Dotter2016mist}) models via \texttt{EXOFASTv2} package \citep{Eastman2017, Eastman2019}.

To derive the stellar parameters of K2-232, we used MESA Isochrones \& Stellar Tracks \citep[MIST;][]{Choi2016mist,Dotter2016mist} models in conjunction with a spectral energy distribution (SED) fit implemented in the \texttt{EXOFASTv2} package \citep{Eastman2017, Eastman2019}. \texttt{EXOFASTv2} combines SED fits to broad-band photometry with isochrone fits to MIST models to constrain the host star's stellar parameters. Our SEDs were constructed from a range of broad-band photometric catalogs: \gaia\, DR3 \citep{GaiaCollaboration2023}, 2MASS \citep{Cutri2003}, and WISE \citep{Cutri2014AllWISE}. See \autoref{tab:toi2005} for all of K2-232's astrometry and photometry used in our SED fits. We adopted Gaussian priors on [Fe/H] and \ensuremath{T_{\rm eff}} derived from \citealt{Brahm2018} which utilized \texttt{ZASPE} \citep{Brahm2015, Brahm2017} synthetic spectral fitting applied to co-added FEROS spectra. We also adopted Gaussian priors on the host star's parallax from \gaia\, DR3 \citep{GaiaCollaboration2023}. We set an upper limit on the $V$-band extinction (\ensuremath{A_{\rm V}}) according to \citealt{Schlafly2011}\footnote{\url{https://irsa.ipac.caltech.edu/applications/DUST/}}, and for all other parameters, we adopted uniform priors. All derived stellar parameters can be found in \autoref{tab:K2-232_exofast.}.

\begin{figure*}
        \centering
        \includegraphics[width=0.8\textwidth]{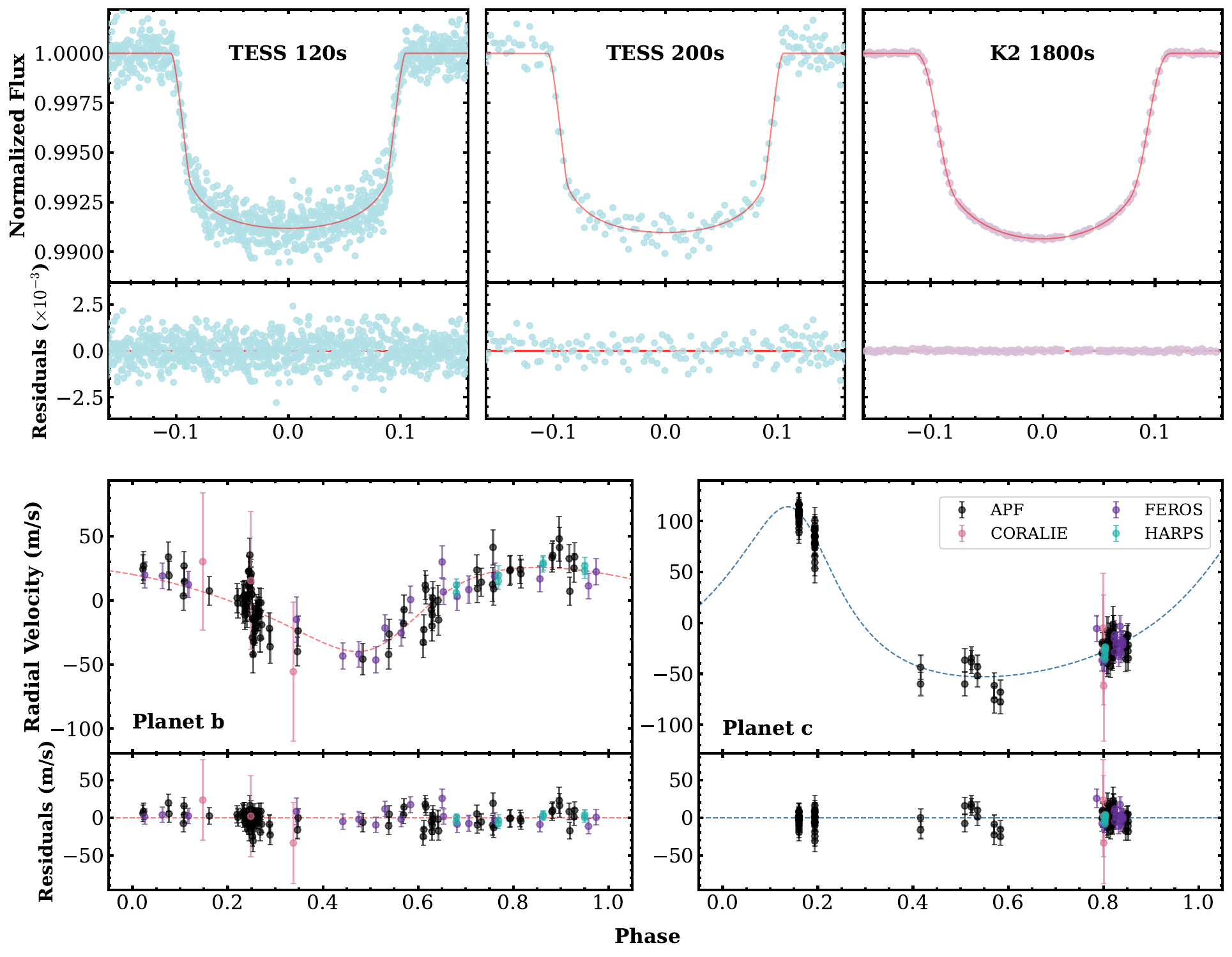}
        \caption{TESS (light blue) and K2 (lavender) flux measurements versus phase for K2-232 showing the transiting warm Jupiter K2-232b, together with APF (black), CORALIE (pink), FEROS (purple), and HARPS (teal) RV measurements versus phase for the K2-232 system. Top panels: TESS and K2 normalized flux measurements phase-folded to planet b's period with 120-second (left), 200-second (center), and 1800-second (right) cadence data. Bottom panels: RV measurements phase-folded to the periods of planet b (left) and planet c (right), with the red and blue dashed lines indicating their respective best-fit curves. All residual data is shown on the bottom of each panel.}
        \label{fig:phaserv}
\end{figure*}

\subsubsection{Acquiring Planetary Parameters}

To derive the planetary parameters of K2-232b and its long-period companion K2-232c, we further utilized the program \texttt{EXOFASTv2}. \texttt{EXOFASTv2} is an exoplanet modeling suite that uses a Differential Evolution Markov Chain Monte Carlo method to fit and characterize exoplanet systems. In this work, photometric data from \ktwo\, and \tess\,-SPOC were jointly analyzed with RVs from the APF \citep{Vogt2014}, CORALIE \citep{CORALIE}, FEROS \citep{1999Msngr..95....8K}, and HARPS \citep{Mayor2003} to provide the best-fit parameters for planets b and c in the K2-232 system. 

Adopting Gaussian priors on the RV offset ($\gamma _{rel}$) for each RV instrument derived from \citealt{Brahm2018}, along with adopting uniform priors for all other planet b and c parameters, we performed a global fit using MIST SEDs, \ktwo\, and \tess\, photometric data, as well as all available RV data for the system. We also adopted quadratic limb darkening coefficients by applying the derived stellar parameters to the Quadratic Limb Darkening applet\footnote[2]{\url{https://astroutils.astronomy.osu.edu/exofast/limbdark.shtml}}.

% \clearpage

For each joint fit, \texttt{EXOFASTv2} implements an automatic burn-in process where all chains must have a $\chi^2$ value below the median $\chi^2$ or everything is discarded \citep{Eastman2013}. The remaining values are then concatenated into the final parameter posteriors. To determine a converged result of our fit we used the recommended convergence criteria by \citealt{Eastman2019} of a Gelman–Rubin statistic <1.01 \citep{Gelman1992} and the number of independent draws from the posterior being >1000. In our fit, meeting the convergence criteria required 400,000 steps. The best-ﬁt models for our phase-folded RVs and transits are shown in \autoref{fig:phaserv}. For planet b, we calculated a best-fit orbital period of $P = 11.1684377 \pm0.0000010$ days, an eccentricity of $e=0.245 ^{+0.023}_{-0.024}$, and a mass of $M = 0.427 ^{+0.039}_{-0.036} \mj$, consistent with literature values. For planet c, we calculated a best-fit orbital period of $P = 1950 ^{+140}_{-120}$ days, an eccentricity of $e=0.352 ^{+0.095}_{-0.076}$, and a minimum mass of $M\sin{i} = 5.31 ^{+0.48}_{-0.45} \mj$. The remaining derived planetary parameters of both planets b and c are listed in \autoref{tab:K2-232_exofast.}. 
 
% The converged fit indicates that, for the inner planet, the orbital period is $11.1684370^{+0.0000032}_{-0.0000033}$ days, the planetary mass is $0.390^{+0.035}_{-0.034}$ $M_J$, and an eccentricity of $0.263^{+0.023}_{-0.024}$. As for the outer companion,  we found a period of $1910^{+140}_{-110}$ days, a minimum mass of $5.12^{+0.49}_{-0.46}$ $M_J$, and an eccentricity of $0.349^{+0.10}_{-0.084}$.

\section{Discussion}
\label{sec:disc}

\subsection{Investigating Outer Companion Eccentricity Distributions}

Several lines of evidence support the theory that hot Jupiters experience dynamically hotter evolution than warm Jupiters. In single-star systems, RM measurements demonstrate that hot Jupiters tend toward spin-orbit misalignment around hot stars \citep{Winn2010, Schlaufman2010, Albrecht2012} while warm Jupiters remain aligned regardless of stellar effective temperature \citep{Rice2022WJs_Aligned, WangXY2024, Espinoza-Retamal2025}. Additionally, hot Jupiters are significantly less likely to host nearby companions than warm Jupiters \citep{Huang2015, Hord2021,Wu2023HJsNotAlone} although some exceptions exist (WASP-47, \citealt{Becker2015}, and Kepler-730, \citealt{Canas2019}). Together, these properties suggest that some form of high-eccentricity migration plays a dominant role in hot Jupiter formation (see, e.g., \citealt{Fabrycky2009,Mustill2015}). However, \citealt{Ngo2016} determined that binary stars are insufficient to serve as the primary hot Jupiter production catalyst.

%\vspace{1cm}
%\clearpage
%\clearpage

Cold Jupiters are prevalent in both hot (51\%, \citealt{Knutson2014}) and warm ($\sim$70\%, \citealt{Bryan2016}) Jupiter systems (see also \citealt{Zink2023}). This could indicate planet-planet interactions are the primary driver of hot Jupiter production rather than stellar companion interactions (e.g., see the K2-290 system; \citealt{Hjorth2019, Best2022}). Consequently, it is expected that the long-term giant companions to hot Jupiters would be more eccentric than those to warm Jupiters, as \citealt{Wu2023HJsNotAlone} predicted in their migration models. As such, and using the California Legacy Survey \citep{Rosenthal2021}, \citealt{Zink2023} found that cold Jupiters hosting inner hot Jupiters are more eccentric than those not.

\begin{figure*}[ht!]
        \centering
        \includegraphics[width=\textwidth]{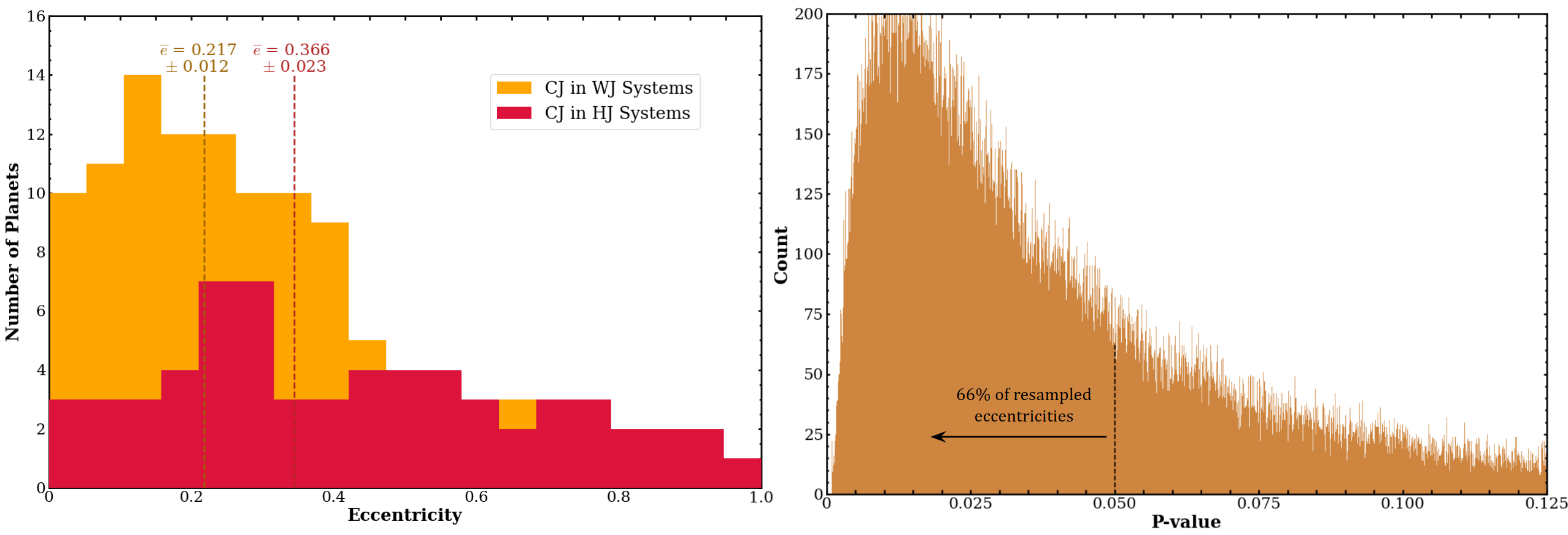}
        \caption{Left panel: Histogram of cold Jupiter orbital eccentricities in warm Jupiter (orange) and hot Jupiter (red) systems resampled 100,000 times. The dark orange and dark red vertical dashed lines represent the mean eccentricity for cold Jupiters in warm and hot Jupiter systems, respectively. Right panel: Histogram of the \textit{p}-value distribution calculated from 100,000 iterations of the two-sample permutation-based Anderson-Darling test comparing both eccentricity distributions. The black dashed line corresponds to \textit{p} = 0.05, below which 66\% of our test iterations lie. These results robustly indicate cold Jupiters hosting inner hot Jupiter companions have higher mean orbital eccentricities than when hosting inner warm Jupiter companions.
}
        \label{fig:no2}
\end{figure*}

In this work, we further investigate this phenomenon by comparing all the eccentricities of hot/cold Jupiter systems to warm/cold Jupiter systems. To do so, we searched the planetary systems composite data table (all confirmed planets) on the NASA Exoplanet Archive \citep{ps} for multi-planet systems ($N_P \geq 2$) with a cold Jupiter ($M$ = 0.3-13$\mj$, $a > 1$ AU) present. We then sorted through the systems that met the above conditions to find those with either a hot ($M$ = 0.3-13$\mj$, $a < 0.1$ AU) or warm Jupiter ($M$ = 0.3-13$\mj$, $a$ = 0.1-1 AU) inner companion.

Through this filtering process we found 16 confirmed cold Jupiters in known hot Jupiter systems and 50 in known warm Jupiter systems. After further system-by-system investigation, we dropped both HD 191939f and WASP-8c from our sample as both cold Jupiters are poorly constrained and lack robust eccentricity estimates, reducing our population sample to 15 cold Jupiters in 15 hot Jupiter systems and 49 cold Jupiters in 44 warm Jupiter systems\footnote{HD 184010 has unmeasured eccentricities for planets c and d (both cold Jupiters), so we bound them with the stability relation $(1/\sin i) + 20e - 2.8 < 0$ from \citealt{Teng2022}: within the resampling loop we draw 1{,}000 uniform $\sin i$ orientations per planet and convert each to an eccentricity upper limit. Each planet's eccentricity is then set to the mean of a half-normal with standard deviation equal to one third of that limit, so the limit falls at the $99.7\%$ ($3\sigma$) bound.}. Relevant planetary and system parameters for each of these cold Jupiters can be found in the table in \autoref{appendix}.

To robustly assess the difference between the eccentricity distributions of cold Jupiters in hot Jupiter systems and those in warm Jupiter systems, we adopted two complementary approaches: a two-sample Anderson-Darling (AD; \citealt{Darling, petit}) test and a comparison of the mean eccentricities ($\bar{e}$) of the two samples. To account for the individual eccentricity measurement uncertainties in both analyses, we also adopted a resampling approach. For cold Jupiters with both upper and lower eccentricity uncertainties available, we uniformly resampled their eccentricities within the reported uncertainty bounds, while for those with only an upper eccentricity limit, we resampled uniformly between 0 and the upper limit.

\noindent\uline{Anderson-Darling Test}

For the AD test, we first compared the two groups using the reported mean eccentricity of each cold Jupiter population and obtained a $p$-value of 0.040, below the significance threshold of 0.05, indicating a significant difference between their eccentricity distributions. We then repeated the AD test while accounting for the individual eccentricity uncertainties using the resampling procedure described above. As we are working with small samples ($N<50$), the AD test may rely on approximate or interpolated $p$-values \citep{Scholz01091987, StatisticalAstro}. To address this limitation, we adopted the permutation method described by \citealt{Goyal2025} to obtain more precise $p$-values. Specifically, we generated 100,000 realizations of the two eccentricity samples and performed a permutation-based AD test for each realization. We found that 66\% of the realizations reject the null hypothesis that the two samples are drawn from the same parent eccentricity distribution at the $p<0.05$ level. The right panel of \autoref{fig:no2} shows the resulting distribution of $p$-values.

\noindent\uline{Mean Eccentricity Comparison}\\
\indent As a complementary test, we compared the mean eccentricities of the two groups while accounting for the individual eccentricity uncertainties using the same resampling procedure. For each realization, we drew one eccentricity value for each cold Jupiter and calculated $\bar{e}$ for each group. We repeated this process 100,000 times to construct the distribution of cold Jupiter eccentricities found in both hot and warm Jupiter systems as is shown on the left panel in \autoref{fig:no2} in red and orange respectively. We find that cold Jupiters in hot Jupiter systems tend to have higher eccentricities ($\bar{e} = 0.366\pm0.023$), whereas those in warm Jupiter systems tend to have lower eccentricities ($\bar{e} = 0.217\pm0.012$). The mean eccentricities of the two samples differ at the 5.75$\sigma$ level, further supporting the difference identified by the AD test. We note that our resampling procedure assumes uniform sampling within the reported eccentricity uncertainty bounds (or between zero and the reported upper limit) and therefore may not perfectly reproduce the true eccentricity probability distribution of each individual planet. This assumption could potentially enhance or weaken the apparent difference between the two samples. Nevertheless, the consistency between the resampling results and the AD tests suggests that our conclusion is robust to the reported eccentricity uncertainties.

\begin{figure*}
        \centering
        \includegraphics[width=\textwidth]{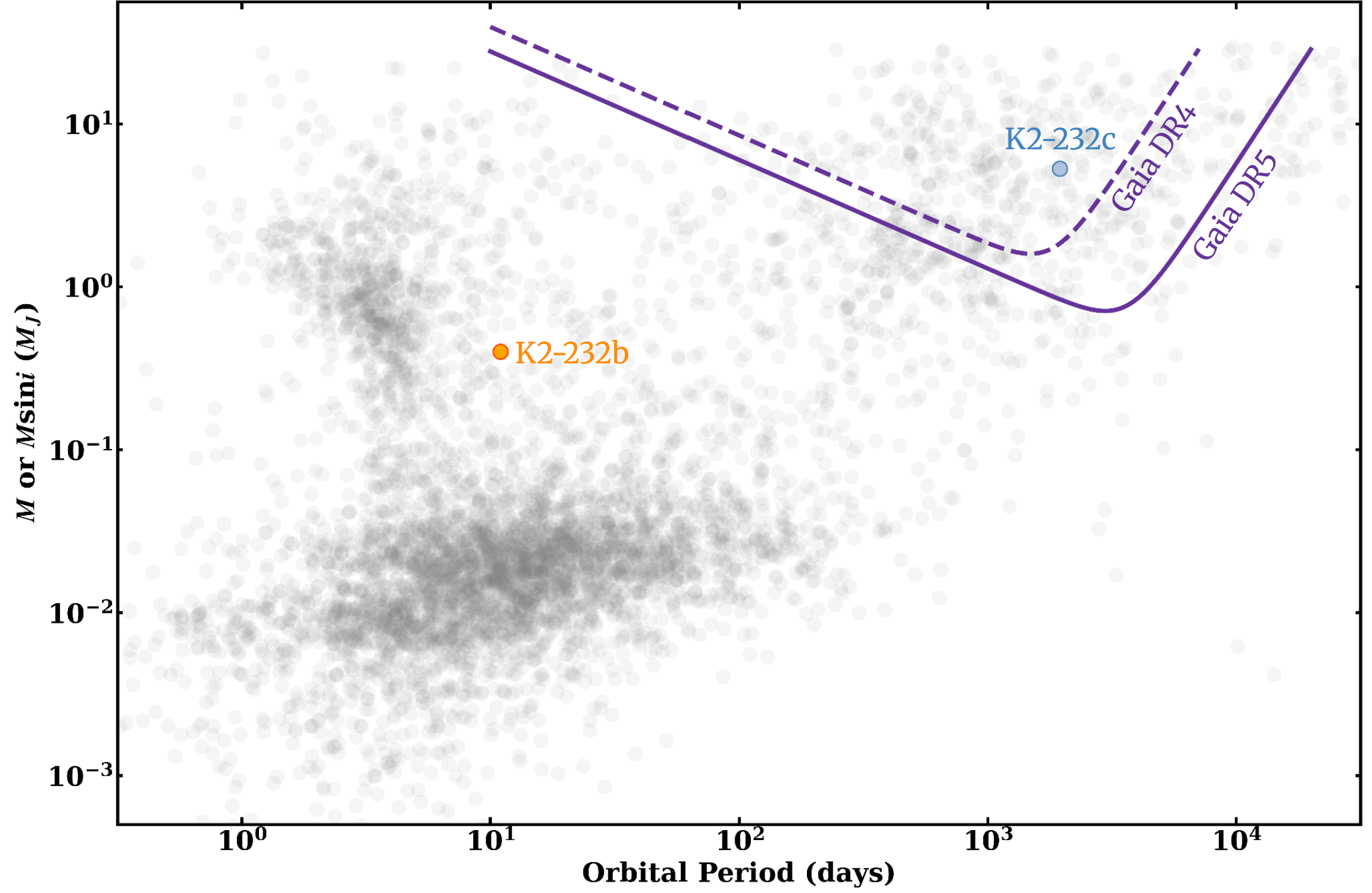}
        \caption{Mass vs. orbital period diagram for entire exoplanet population. Marker transparency values are determined by the distance to the system with fainter points being further away and more opaque points being closer by. The location of K2-232b and K2-232c are labeled on this diagram (in orange and blue respectively). The purple dashed line indicates the detection sensitivity of Gaia DR4 and the purple solid line, for that of DR5. These sensitivities are calculated for a 1 $M_{\odot}$ star viewed from 100 pc with an astrometric precision of 34 $\mu$as \citep[e.g.,][]{Perryman2014} for a 5 and 10 year campaign. K2-232c falls well within the sensitivity region for Gaia DR4 suggesting a likely detection. 
}
        \label{fig:no4}
\end{figure*}

% \textbf{We next investigate the statistical significance of the resampled cold Jupiter eccentricity distributions in the 15 hot Jupiter systems and the 44 warm Jupiters systems. To determine whether they are unique populations or can be explained by the same parent distribution (our null hypothesis), we used the two-sample AD test as it has been demonstrated to be more statistically robust than the standard two-sample KS test, exhibiting enhanced sensitivity near the tails of the cumulative distribution functions.} However, as we are working with a small sample size ($N<50$), the AD test is prone to using approximate or interpolated values \citep{Scholz01091987, StatisticalAstro}. To combat this, we utilized a permutation method described in \citealt{Goyal2025} to increase the resolution of the \textit{p}-value grid. First, we resampled all of our eccentricity values 100,000 times and then ran a permutation-based AD test for each resampling. We recovered evidence in 66\% of our test iterations to reject our null hypothesis of a shared parent eccentricity distribution at the \textit{p} = 0.05 level. The right panel in \autoref{fig:no2} shows the distribution of \textit{p}-values calculated during this procedure. 

Together, these results indicate that cold Jupiters in hot Jupiter systems are significantly more eccentric than those in warm Jupiter systems, consistent with the findings of \citealt{Zink2023}. This contributes an independent line of evidence supporting that hot Jupiters have dynamically hotter histories than warm Jupiters. 

\subsection{Observing K2-232c With \gaia\,}

The current RV sample of cold Jupiter companion systems is not likely to grow much larger in the near future as detecting cold giant companions is observationally prohibitive. These giants have too long of a baseline, their inner companions are intrinsically rare, and they typically don't have bright enough host stars for efficient RV follow-up. Astrometry, however, may soon provide an influx of new cold Jupiter companions.   

The \gaia\, mission is expected to detect at least several thousand exoplanets using the astrometry method, most of them being cold Jupiters \citep{Perryman2014,Lammers2025} and the first of which already being reported using \gaia\, DR3 data \citep{Stefansson2025}. Here, we evaluate the possibility of detecting K2-232c using the full suite of epoch astrometric data from \gaia\, DR4 following the methodology of \citealt{Espinoza-Retamal2023}. Briefly, this methodology involves injection and recovery simulations. We simulated \gaia\, DR4 astrometric data using the posteriors of planet c, and tried to recover the parameters of the system using the Fisher information matrix for a joint model of astrometry and RVs. In 10,000 simulations, and assuming that the inclination of the orbit of planet c is isotropically distributed, we estimated a probability of $\sim$75$\%$ of detecting planet c and measuring the mutual inclination between planet b and c using future \gaia\, DR4 astrometry and current RVs. This relatively high probability results from a combination of facts that produce an astrometric signal large enough to be detected with \gaia\,. These include a bright ($G=9.72$) and nearby ($d\sim130$ pc) star, a long orbital period of planet c, and a long baseline of $\sim$7.5 years of observations (considering both astrometry and RVs) that covers almost one and a half orbits of planet c.

In \autoref{fig:no4}, we plot the current confirmed exoplanet population distribution, highlighting the locations of planets b and c as well as the expected observational sensitivities for both Gaia DR4 and DR5. These calculated sensitivities are regions where the astrometric signal produced by a planet is larger than the error bar produced from Gaia. We assume we should detect all systems where this is true until the orbital period is longer than the observational baseline. 

This plot visually represents the likelihood for observing planet c with Gaia as well as other cold Jupiters similar to it. Falling well within the sensitivity region of Gaia DR4 and combined with our estimated detection probability of $\sim$75$\%$, we are confident Gaia will provide us with important new information about K2-232c that will elucidate both the history and characteristics of the system. This includes, for example, the mutual inclination between the orbits of planets b and c, which can be used as a tracer of the system's dynamical history (see, e.g., \citealt{Espinoza-Retamal2023}).

\begin{acknowledgements}

 We are grateful for helpful discussions with Cristobal Petrovich, Armaan Goyal, Kyle Hixenbaugh, Emma Dugan, Josette Wright, and Jace Rusznak during the preparation of this manuscript. The authors would also like to extend their gratitude toward the anonymous referee, whose comments and suggestions greatly improved the overall clarity of this work. J.A.R. thankfully acknowledges support from the Indiana Space Grant Consortium Fellowship program for partially funding this work. X.Y.W. acknowledges support from the Sullivan Prize Fellowship. J.A.B. acknowledges that part of this research was carried out at the Jet Propulsion Laboratory, California Institute of Technology, under contract with the National Aeronautics and Space Administration (NASA). J.I.E.-R. gratefully acknowledges support from the John and A-Lan Reynolds Faculty Research Fund, from the ANID BASAL project FB210003, and from the ANID Doctorado Nacional grant 2021-21212378. This work was supported in part by the NASA Exoplanets Research Program  NNH23ZDA001N-XRP (Grant No. 80NSSC24K0153), the NASA TESS General Investigator Program, Cycle~7, NNNH23ZDA001N-TESS (Grant No. 80NSSC25K7912), and the Heising-Simons Foundation (Grant \#2023-4050). We acknowledge funding support from grant JWST-GO-09025.010-A provided by NASA via the Space Telescope Science Institute under the JWST General Observers Program \#9025. The Space Telescope Science Institute is operated by the Association of Universities for Research in Astronomy, Inc., under NASA contract NAS 5-03127 for JWST. S. W. gratefully acknowledges support from the John and A-Lan Reynolds Faculty Research Fund.

\end{acknowledgements}

% \clearpage

\appendix
\section{Additional Tables}
\label{appendix}

\setcounter{table}{0}
\renewcommand{\thetable}{A\arabic{table}}
\startlongtable
\begin{deluxetable}{ccccc}
\label{tab:appen}
%[h]
% \tabletypesize{\tiny}
\tablehead{
\multicolumn{5}{c}{\textbf{Cold Jupiters in Hot Jupiter Systems}} \\
\colhead{{\textbf{Planet Name}}} & \colhead{{\textbf{Number of Planets}}} & \colhead{{\textbf{Planet Mass (\mj)}}} & \colhead{{\textbf{Semi-Major Axis (AU)}}} & \colhead{{\textbf{Eccentricity}}}
}
\startdata
HIP 14810 d & 3 & $0.59\pm0.10$ & $1.94\pm0.13$ & $0.185\pm0.035$ \\
WASP-47 c & 4 & $1.406419^{+0.06}_{-0.07}$ & $1.393\pm0.014$ & $0.264^{+0.007}_{-0.012}$ \\
HD 83443 c & 2 & $1.5^{+0.5}_{-0.2}$ & $8.0\pm0.8$ & $0.760^{+0.046}_{-0.047}$\\
HD 187123 c & 2 &  $2.87^{+0.40}_{-0.42}$ & $4.89\pm0.53$ & $0.252\pm0.033$\\
WASP-41 c & 2 & $3.18\pm0.20$ & $1.07\pm0.03$ & $0.294\pm0.024$\\
HAT-P-17 c & 2 & $3.4^{+1.1}_{-0.7}$ & $5.6^{+3.5}_{-1.4}$ & $0.39^{+0.23}_{-0.17}$\\
KELT-6 c & 2 & $3.71\pm0.21$ & $2.39\pm0.11$ & $0.210^{+0.039}_{-0.036}$\\
HAT-P-44 c & 2 & $4.0^{+1.4}_{-0.8}$ & $1.752\pm0.025$ & $0.494\pm0.081$\\
HD 217107 c & 2 & $4.37^{+0.13}_{-0.10}$ & $5.922^{+0.035}_{-0.044}$ & $0.3918^{+0.0064}_{-0.0067}$\\
WASP-132 d & 3 & $5.16\pm0.52$ & $2.71\pm0.12$ & $0.120\pm0.078$\\
Pr0211 c & 2 & $7.79\pm0.33$ & $5.5^{+3.0}_{-1.4}$ & $0.71\pm0.11$\\
ups And d & 3 & $10.25^{+0.70}_{-3.30}$ & $2.51329\pm0.00075$ & $0.2987\pm0.0072$\\
HAT-P-2 c & 2 & $10.7^{+5.2}_{-2.2}$ & 8.968117* & $0.37^{+0.13}_{-0.12}$\\
HD 118203 c & 2 & $11.79^{+0.69}_{-0.63}$ & $6.28^{+0.10}_{-0.11}$ & $0.26^{+0.03}_{-0.02}$\\
HATS-59 c & 2 & $12.70\pm0.87$ & $2.504\pm0.035$ & <0.083\\
\hline
\multicolumn{5}{c}{\textbf{Cold Jupiters in Warm Jupiter Systems}} \\
\hline
HD 184010 c & 3 & $0.30^{+0.03}_{-0.06}$ & $1.334^{+0.013}_{-0.005}$ & $0^{\star}$\\
HD 184010 d & 3 & $0.45^{+0.04}_{-0.06}$ & $1.920\pm0.012$ & $0^{\star}$\\
HIP 57274 d & 3 & $0.5267\pm0.03$ & $1.01$ & $0.27\pm0.05$\\
HIP 14810 d & 3 & $0.59\pm0.10$ & $1.94\pm0.13$ & $0.185\pm0.035$\\
HD 37124 c & 3 & $0.652\pm0.052$ & $1.7100\pm0.0065$ & $0.125\pm0.055$\\
HD 141399 e & 4 & $0.66\pm0.10$ & $5.0\pm1.5$ & $0.26\pm0.22$\\
HD 37124 d & 3 & $0.696\pm0.059$ & $2.807\pm0.038$ & $0.16\pm0.14$\\
HD 207832 c & 2 & $0.73^{+0.18}_{-0.05}$ & $2.112^{+0.087}_{-0.045}$ & $0.27^{+0.22}_{-0.10}$\\
HD 134987 c & 2 & $0.82\pm0.03$ & $5.8\pm0.5$ & $0.12\pm0.02$\\
HD 155358 c & 2 & $0.82\pm0.07$ & $1.02\pm0.02$ & $0.16\pm0.1$\\
HD 141399 d & 4 & $1.18\pm0.08$ & $2.09\pm0.06$ & $0.074\pm0.025$\\
Kepler-451 c & 3 & $1.61\pm0.14$ & $1.972970^*$ & $0.29\pm0.07$\\
HD 82943 b & 2 & $1.681\pm0.028$ & $1.18306\pm0.00062$ & $0.162\pm0.036$\\
HD 12661 c & 2 & $2.83^{+0.82}_{-0.67}$ & $2.8145\pm0.0238$ & $0.031\pm0.022$\\
HD 163607 c & 2 & $2.201\pm0.037$ & $2.39\pm0.11$ & $0.080\pm0.014$\\
HD 73526 c & 2 & $2.25\pm0.13$ & $1.03\pm0.02$ & $0.28\pm0.05$\\
HD 147873 c & 2 & $2.30\pm0.18$ & $1.36\pm0.05$ & $0.23\pm0.03$\\
Kepler-539 c & 2 & $2.4\pm1.2$ & $2.42^{+0.50}_{-0.51}$ & $0.5\pm0.1$\\
Kepler-432 c & 2 & $2.43^{+0.22}_{-0.24}$ & $1.178029^*$ & $0.498^{+0.029}_{-0.059}$\\
HD 60532 c & 2 & $2.51\pm0.16$ & $1.60\pm0.04$ & $0.03\pm0.02$\\
HD 73344 d & 3 & $2.55^{+0.56}_{-0.46}$ & $6.70^{+0.25}_{-0.26}$ & $0.18^{+0.14}_{-0.12}$\\
Kepler-88 d & 3 & $3.1\pm0.2$ & $2.466329866$ & $0.41\pm0.03$\\
HD 37605 c & 2 & $3.19\pm0.38$ & $3.74\pm0.21$ & $0.030\pm0.012$\\
HD 92788 c & 2 & $3.67^{+0.30}_{-0.25}$ & $10.50^{+2.90}_{-0.55}$ & $0.46^{+0.12}_{-0.03}$\\
55 Cnc d & 5 & $3.878\pm0.068$ & $ 5.60\pm0.10$ & $0.0913\pm0.0067$\\
HD 160691 b & 4 & $<4.3$ & $1.50\pm0.02$ & $0.0505^{+0.0094}_{-0.0093}$\\
HD 160691 c & 4 & $<4.4$ & $4.17\pm0.07$ & $0.039\pm0.011$\\
HD 11506 b & 3 & $4.880^{+1.986}_{-0.333}$ & $2.885\pm0.016$ & $0.379\pm0.009$\\
TIC 139270665 c & 2 & $4.89^{+0.66}_{-0.37}$ & $2.00^{+0.93}_{-0.31}$ & $0.566^{+0.120}_{-0.069}$\\
HD 13908 c & 2 & $5.13\pm0.25$ & $2.03\pm0.04$ & $0.12\pm0.02$\\
HD 148164 c & 2 & $5.16\pm0.82$ & $6.15\pm0.50$ & $0.125\pm0.017$\\
Kepler-56 d & 3 & $5.61\pm0.38$ & $2.16\pm0.08$ & $0.20\pm0.01$\\
TOI-2295 c & 2 & $5.61^{+0.23}_{-0.24}$ & $2.018^{+0.040}_{-0.042}$ & $0.194\pm0.012$\\
TOI-4562 c & 2 & $5.77^{+0.37}_{-0.56}$ & $5.219\pm0.002$ & $0.122^{+0.027}_{-0.026}$\\
HD 27894 d & 3 & $6.493^{+0.987}_{-0.353}$ & $5.362^{+0.206}_{-0.223}$ & $0.343^{+0.031}_{-0.026}$\\
HD 147018 c & 2 & $6.56\pm0.32$ & $1.922\pm0.039$ & $0.133\pm0.011$\\ 
HIP 67851 c & 2 & $6.937^{+2.045}_{-0.518}$ & $4.55\pm0.05$ & $0.30\pm0.03$\\
TOI-2537 c & 2 & $7.23^{+0.52}_{-0.45}$ & $2.78^{+0.22}_{-0.15}$ & $0.287^{+0.060}_{-0.052}$\\
Kepler-419 c & 2 & $7.3\pm0.4$ & $1.68\pm0.03$ & $0.184\pm0.002$\\
HD 169830 c & 2 & $7.6690^{+1.9370}_{-2.7550}$ & $3.075^{+0.132}_{-0.146}$ & $0.246^{+0.022}_{-0.018}$\\
HIP 8541 b & 2 & $7.714^{+1.895}_{-0.103}$ & $3.729^{+0.300}_{-0.168}$ & $0.360^{+0.060}_{-0.047}$\\
HD 43197 c & 2 & $7.868^{+1.760}_{-1.599}$ & $8.5400^{+2.3340}_{-1.5840}$ & $0.1490^{+0.1120}_{-0.0870}$\\
K2-99 c & 2 & $8.4\pm0.2$ & $1.43\pm0.01$ & $0.210\pm0.009$\\
HD 74156 c & 2 & $8.665^{+1.385}_{-0.470}$ & $3.678^{+0.145}_{-0.159}$ & $0.377\pm0.006$\\
TOI-4600 c & 2 & $<9.27$ & $1.152\pm0.068$ & $0.21^{+0.29}_{-0.14}$\\
HD 156279 c & 2 & $9.7500^{+1.3190}_{-0.6050}$ & $5.486^{+0.219}_{-0.240}$ & $0.261\pm0.006$\\
TYC 1422-614-1 c & 2 & $10\pm1$ & $1.37\pm0.06$ & $0.048^{+0.020}_{-0.014}$\\
HD 38529 c & 2 & $10.380^{+1.025}_{-0.884}$ & $3.226^{+0.131}_{-0.144}$ & $0.357\pm0.005$\\
HD 11506 d & 3 & $12.8^{+0.6}_{-0.5}$ & $18.20^{+0.06}_{-0.09}$ & $0.29^{+0.02}_{-0.03}$\\
\enddata
\tablenotetext{}{Note: *Semi-major axes calculated when only period values are provided in the literature.
$^{\star}$Eccentricity was selected based on a model, not directly measured.}
\end{deluxetable}

% \clearpage

\bibliography{main}{}
\bibliographystyle{aasjournal}

%% This command is needed to show the entire author+affilation list when
%% the collaboration and author truncation commands are used.  It has to
%% go at the end of the manuscript.
%\allauthors

%% Include this line if you are using the \added, \replaced, \deleted
%% commands to see a summary list of all changes at the end of the article.
%\listofchanges

\end{document}